\documentclass[11pt,a4paper]{article}

\usepackage[toc,page]{appendix}
\usepackage{authblk}
\usepackage{genres}

\usepackage{color}
\usepackage{fullpage}
\usepackage{amsmath,bm}
\usepackage{amsfonts}
\usepackage{graphicx}
\usepackage{paralist}

\usepackage{algorithm}		% for algorithm
\usepackage{algpseudocode}	% for algorithm

\begin{document}

\title{Statistical modeling of isoform splicing dynamics from RNA-seq time series data}

\author{Yuanhua Huang\,$^{1}$ and Guido Sanguinetti\,$^{1,2}$
\footnote{To whom correspondence should be addressed.
Tel: +44 131 6505136; Fax: +44 131 6511426; Email: G.Sanguinetti@ed.ac.uk}}

\affil{
$^{1}$School of Informatics, University of Edinburgh, Edinburgh EH8 9AB, UK
and
$^{2}$Centre for Synthetic and Systems Biology (SynthSys), University of Edinburgh, Edinburgh, EH9 3BF, UK\vspace*{-18pt}}
\date{}
\maketitle

\begin{abstract}
Isoform quantification is an important goal of RNA-seq experiments, yet it remains problematic for genes with low expression or several isoforms. These difficulties may in principle be ameliorated by exploiting correlated experimental designs, such as time series or dosage response experiments. Time series RNA-seq experiments, in particular, are becoming increasingly popular, yet there are no methods that explicitly leverage the experimental design to improve isoform quantification. Here we present DICEseq, the first isoform quantification method tailored to correlated RNA-seq experiments. DICEseq explicitly models the correlations between different RNA-seq experiments to aid the quantification of isoforms across experiments. Numerical experiments on simulated data sets show that DICEseq yields more accurate results than state-of-the-art methods, an advantage that can become considerable at low coverage levels. On real data sets, our results show that DICEseq provides substantially more reproducible and robust quantifications, increasing the correlation of estimates from replicate data sets by up to 10\% on genes with low or moderate expression levels (bottom third of all genes). Furthermore, DICEseq permits to quantify the trade-off between temporal sampling of RNA and depth of sequencing, frequently an important choice when planning experiments. Our results have strong implications for the design of RNA-seq experiments, and offer a novel tool for improved analysis of such data sets. 
Python code is freely available at \texttt{http://diceseq.sf.net}. 
\end{abstract}

\section{Introduction}
In most eukaryotes, alternative splicing is an important post-transcriptional mechanism of regulation of gene expression, and largely increases the diversity of the proteome \citep{Graveley2001}. For example, over 90\% of human genes have multiple isoforms \citep{Wang2008}. Several lines of evidence indicate that alternative splicing plays a vital role in regulating biological processes \citep{Blencowe2006}, and its failure often causes serious diseases \citep{Scotti2016}.%\citep{Venables2009,Scotti2016}. 
The study of splicing has been revolutionised by the advent of  high-throughput transcriptome sequencing (RNA-seq) techniques which enable unbiased sampling of the transcriptome and have greatly contributed to uncover novel biological functions for alternative splicing \citep{Wang2009}. More recently, RNA-seq technologies have been combined with biotin labelling treatment to provide kinetic measurements of RNA transcription and splicing with high temporal resolution \citep{Windhager2012, Eser2015, Barrass2015}, providing invaluable mechanistic insights in the dynamics of splicing.

At the current stage of the technology, sequenced reads in RNA-seq experiments are much shorter than almost all eukaryotic transcripts. Thus, most reads from an RNA-seq experiment cannot be unambiguously aligned to a specific isoform. While in some cases a high level of coverage may obviate the problems, in many cases the number of reads that map to a single isoform is too low; when many isoforms are present, there may be no unambiguously assigned reads. To address this problem, several probabilistic methods were proposed to quantify the isoform proportion, for example IsoEM \citep{Nicolae2011}, Cufflinks \citep{Trapnell2010}, MISO \citep{Katz2010}, and BitSeq \citep{Glaus2012}. All of these methods introduce latent variables to model the {\it identity} of a read, i.e. which isoform it came from, and then reconstruct isoform proportions by maximum likelihood or by computing a posterior distribution from the observed read distribution.

Most of these computational methods can quantify the isoform proportions accurately in many cases \citep{Kanitz2015}, %\citep{Steijger2013,Kanitz2015}, 
however for all methods isoform quantification at low coverages remains challenging. A natural approach in these cases is to exploit additional information, for example exploiting correlations across different experiments arising out of structured experimental designs such as time series or dosage response experiments. Time series RNA-seq designs, in particular, are becoming increasingly popular as an effective tool to investigate the dynamics of gene expression in a range of systems \citep{Bar2012,Tuomela2012,Zhang2014,Honkela2015}. To our knowledge, no methods have been proposed that can exploit structured experimental designs in order to improve isoform estimation. This methodological gap also negatively affects the ability to design effectively experiments: for example, it is difficult to understand whether resources should be invested in gathering more time points, or in sequencing at a deeper level a more limited number of samples.

In this article, we present a new methodology, DICEseq (Dynamic Isoform spliCing Estimator via sequencing data) to jointly estimate the dynamics of isoform proportions from RNA-seq experiments with structured experimental designs. DICEseq is a Bayesian method based on a mixture model whose mixing proportions  represent isoform ratios, as in \citep{Katz2010,Glaus2012}; however, DICEseq incorporates the correlations induced by the structured design by coupling the isoform proportions in different samples through a latent Gaussian process (GP). By doing so, DICEseq effectively transfers information between samples, borrowing strength which can aid to identify the isoform proportions. Our results show that DICEseq consistently improves in accuracy and reproducibility over the state of the art. This improvement can be very significant for a large fraction of genes: on one real data set, the correlation between estimates from replicate data sets increased by over 10\% across one third of the genes as a result of taking temporal information into account. Furthermore, simulation studies indicate that DICEseq can be an important tool in experimental design, enabling an effective trade-off of resources between sequencing depth and sample numbers. DICEseq therefore offers an effective way to maximise information extraction from complex high-throughput data sets.

%\begin{figure*}[!tpb]
%\centerline{\includegraphics[width=0.94\textwidth]{figure/Figure1.png}}
%\caption{A cartoon comparison between separate and joint analysis of time-series RNA-seq experiments.}
%\end{figure*}

\section{Methods}
\subsection{Mixture modelling of RNA-seq data}
We briefly review here the mixture modelling framework for isoform identification (MISO), as described in \citep{Katz2010}. We will describe the model on a per gene basis; the output of an RNA-seq experiment is therefore $N$ reads $R_{1:N}$ aligned to a gene with $C$ isoforms. Each read $R_n$ has its {\it identity} $I_n\in\{1,\ldots,C\}$, i.e. which specific isoform it originated from, but, unless the read is aligned to isoform specific region, e.g., a junction,
we will not know its identity. The proportion of each specific isoform within the pool of total mRNA is defined by the vector $\Psi$, whose entries must be positive and  sum to 1. We can then  define the likelihood of isoform proportions $\Psi$ as mixture model as follows
\begin{equation}
P(R_{1:N}|\Psi) = \prod_{n=1}^{N} \sum_{I_n=c}^C  P(R_{n}|I_n) P(I_n| \Psi).
\end{equation}
The conditional distribution of $I_n| \Psi$ is assumed to be Multinomial,  $(I_n| \Psi) \sim \mathtt{Multinomial} (\Psi * w)$ where $w$ is a weight vector adjusting the isoform proportion by the effective length of each isoform. The model is completed by specifying a prior distribution over the isoform proportion vector $\Psi$, which in \citep{Katz2010} was chosen to be a Dirichlet on $\Psi$. Extending the MISO model to time series RNA-seq experiments involves a choice on how to model temporal correlations between the values of $\Psi$ at different time points; we will use a flexible non-parametric prior in the form of a Gaussian process for this.

\subsection{Gaussian processes}
Gaussian processes (GPs) are a generalisation of the multivariate normal distribution to infinite-dimensional random functions. The key property of a GP is that all of its finite dimensional marginals are multivariate normals; in other words, evaluating a random function drawn from a GP at a finite set of points yields a normally distributed random vector. A GP over a suitable input space $\mathcal{T}$ is uniquely specified by a mean function $m\colon\mathcal{T}\rightarrow\mathbb{R}$ and a covariance function $k\colon\mathcal{T}\times\mathcal{T}\rightarrow\mathbb{R}$, which models how correlations between function outputs depend on the inputs. In this paper, we will identify the input space $\mathcal{T}$ with the time axis, and use as a covariance function the {\it squared exponential} (or RBF) covariance
\begin{equation}
k(t_1, t_2) = \theta_1 \mathtt{exp}(-\frac{1}{2\theta_2}(t_1 - t_2)^2).\label{RBFcov}
\end{equation}
The covariance function depends on two hyper-parameters, the prior variance $\theta_1$ and the (squared) correlation lengthscale $\theta_2$.

The fundamental property of GPs relates the abstract function space view of GPs reported above with the explicit parametric form of their finite dimensional marginals. Let $f$ denote a random function sampled from a GP, $(t_1,\ldots,t_N)$ denote a set of input (time) points and $\mathbf{f}=(f(t_1),\ldots,f(t_N))$ the vector obtained by evaluating the function $f$ over the input points. Then, we have that

\begin{equation}
f\sim\mathcal{GP}(m,k)\leftrightarrow \mathbf{f}\sim \mathcal N(\bm m, K)\label{GPmarginals}
\end{equation}
where $\bm m$ and $K$ are obtained by evaluating the mean and covariance functions over the set of points  $(t_1,\ldots,t_N)$ (and pairs thereof). The fundamental property \eqref{GPmarginals} is key to the success of GPs as a practical tool for Bayesian inference: given observations of the function values $\mathbf{y}$, it is in principle straightforward to obtain posterior predictions of the function values everywhere by applying Bayes' theorem
\begin{equation}
p(f(t_{new}\vert\mathbf{y}))\propto\int d\mathbf{f} p(\mathbf{f},f(t_{new}))p(\mathbf{y}\vert\mathbf{f})\label{GPpred}
\end{equation}
If the observation noise model $p(\mathbf{y}\vert\mathbf{f})$ is Gaussian, then the integral in \eqref{GPpred} is analytically computable. Notice that equation \eqref{GPpred} provides a way of predicting the latent function {\it at all time points}, not just the observation points. In the following, we describe an algorithm to approximate the computation of \eqref{GPpred} for multinomial observations. For a thorough review of GPs and their use in modern machine learning, we refer the reader to the excellent book \citep{Rasmussen2006}. 

\subsection{Posterior of splicing dynamics with GP prior}\label{MCMCalgorithm}
Given a set of RNA-seq reads $\bm R=[R_{1:N_1}^{(1)},...,R_{1:N_T}^{(T)}]$ for $T$ time points that are aligned to a gene with $C$ isoforms, the posterior of the splicing dynamics for the isoform proportions $\bm\Psi =[\Psi_{1:C}^{(1)},...,\Psi_{1:C}^{(T)}]$ is as follows,
\begingroup\makeatletter\def\f@size{9}\check@mathfonts
\begin{equation}
\begin{split}
& P(\bm\Psi\vert \bm\Theta,\bm R) \propto P(\bm\Theta) P(\bm\Psi|\bm\Theta) \times \prod_{t=1}^{T} P(R_{1:N_t}^{(t)}|\Psi^{(t)})\\
& \propto P(\bm\Theta) P(\bm\Psi|\bm\Theta) \times \prod_{t=1}^{T} \prod_{n=1}^{N_t} \sum_{I_n^{(t)}=c}^C  P(R_{n}^{(t)}|I_n^{(t)}) P(I_n^{(t)}| \Psi^{(t)})\\
\end{split}
\end{equation}
\endgroup

We assume that $\Psi$ is a $\mathtt{Softmax}$ function of latent variable $Y$, i.e., $\psi_c = e^{y_c} / \sum_{i=1}^C e^{y_i}$, and $y_C=0$ to make the correspondence. Also $Y_c=[y_c^{(1)},...,y_c^{(T)}]$ follows a Gaussian process with its isoform specific hyperparameters $\bm\theta_c$ and mean $\bm m_c$.

We assume in the following that the prior GP has zero mean, but this can be adjusted in a straightforward way to a more informative prior. Hyperparameters can also be sampled, however this leads to a much more complex inference problem since latent function values and hyperparameters are strongly correlated. We therefore fix $\theta_{c,1}=3.0$, so that the 95\% prior confidence intervals of $\psi$ at an independent time point goes from 0.03 to 0.97, and set the second hyperparameter $\theta_2$ empirically to account for approximately 20-40\% of the duration of the experiment. A sensitivity analysis to $\theta_2$ is provided in Supplementary Table S1.

Having defined the posterior of the splicing dynamics, we introduce a Metropolis-Hasting sampler in Algorithm 1, which is a Markov chain Monte Carlo (MCMC) method, to infer the posterior of the splicing dynamics.

\begin{algorithm}
  \caption{Metropolis-Hastings sampler for posterior of latent $\bm Y$}
  \begin{algorithmic}
  	\State {\textbf{Require:} $T, \bm R, \bm\Theta, \lambda$}
    \State {\textbf{Initialize:} $\bm Y^{(0)}$}
    \State {\textbf{Calculate:} $\bm\Psi^{(0)}=\mathtt{Softmax}(\bm Y^{(0)})$;
    $\bm K = \mathtt{GPcov}(\Theta, T)$}
    \For{$i=0$ to $H$}
		    \State {\textbf{Sample:} $\mu \sim U(0,1)$}
    		\State {\textbf{Sample:} $\bm Y^* \sim Q_y(\bm Y^*|\bm Y^{(i)}, \lambda\bm K)$}
    		\State {\textbf{Calculate:} $\bm\Psi^*=\mathtt{Softmax}(\bm Y^*)$}
			\If{$\mu < \mathtt{min} \Big\{ \dfrac 
			{P(\bm\Psi^*|\bm R) \times
			 Q_y(\bm Y^{(i)}| \bm Y^*, \lambda\bm K)}
			{P(\bm\Psi^{(i)}|\bm R)  \times
			 Q_y(\bm Y^*|\bm Y^{(i)},\lambda\bm K)}, 1 \Big\}$}
    			\State {$\bm Y^{(i+1)} \leftarrow \bm Y^*$; 
    					$\bm \Psi^{(i+1)} \leftarrow \bm \Psi^*$}
  			\Else
    			\State {$\bm Y^{(i+1)} \leftarrow \bm Y^{(i)}$; 
    					$\bm \Psi^{(i+1)} \leftarrow \bm \Psi^{(i)}$}
  			\EndIf
    	\EndFor
  \end{algorithmic}
\end{algorithm}

Here, the proposal distribution $Q_y$ for $Y_c$ is a multivariate Gaussian distribution, whose mean is the last accepted $Y_c^{(i)}$, and the covariance matrix is defined by the fixed hyper-parameters $\bm\theta_c$ and the times $T$, but adjusted to the data itself, including the empirical variance of $y$, the number of isoforms, and number of time points, to ensure the 30-50\% acceptance ratio. Namely, $\hat K_c=\lambda K_c; \lambda=(5 \sigma_y^2)/(C T \theta_{c,1})$, and the proposal distribution is $\mathcal{N}(Y_c^{(i)}, \hat K_c)$. Notice that, in contrast to the MISO algorithm \citep{Katz2010}, our sampler directly collapses the read identity variables, leading to considerable speedups when the number of isoforms is not too high.

For each gene, the initial MCMC chain contains $1000$ iterations. Then the Geweke diagnostic $Z$ score %\citep{Geweke1991} 
is applied to check the convergence of $\bm Y$, using the first 10\% and the last 50\% iteration of the sampled chain. If $|Z| > 2$, then 100 more iterations will be added until the criterion is passed.

\subsection{Reads probability and bias correction}
DICEseq supports both single-end and paired-end reads. Here we describe the situation of paired-end reads; for single-end reads, just change the fragment length into read length. Given a read (pair) $R_n$ mapping to an isoform $c$, the reads probability $P(R_n|I_n=c)$ could be defined by taking information of the fragment length $l_f$, the alignment quality $\mathtt{mapq}$, and the reads position $p$, as follow,
\begin{equation}
P(R_n|I_n)=P(l_f|I_n)P(p|I_n,l_f)P(R_n|\mathtt{mapq})
\end{equation}

Here, we apply a Gaussian distribution to model the distribution of fragment length. The parameters (mean and variance) could be either set by user or learnt from the data itself. In some species, most reads can be very well mapped to a single position of the genome, and we could simply use uniquely mapped reads. However, in some other species, such as yeast, which contains many paralogs, there are higher chances to align a read to multiple positions. In the latter case for keeping multiply aligned reads, the $\mathtt{mapq}$ score will be taken into account, as $P(R_n|\mathtt{mapq})=1-10^{-\mathtt{MAPQ}/10}$, and we take the score of the better aligned mate for reads pair.

The reads position could be assumed to come from a uniform distribution, or could explicitly model sequence and position biases. In both cases, we could describe the probability as follows,
\begin{equation}
P(p|I_n=c,l_f)=\frac {b_c(p)} {\sum_{j=1} ^{l_k-l_f+1} {b_c(j)}}
\end{equation} 
where $b_c(p)$ is relative weight of a position $p$. For uniform distribution, $b_c(p) \equiv 1$, so that $P(p|I_n,l_f)=1/{l_k-l_f+1}$. For the bias distribution, we employed the bias correction model that was proposed by Roberts \emph{et al} \citep{Roberts2011} to correct the position and sequence bias.

Briefly, Roberts \emph{et al}'s model of position bias tries to estimate which fractional position is preferred for sequencing. Thus, 20 bins from the beginning to the end of the isoform were used to count aligned reads, and isoforms are also divided into 5 groups based on their length. The sequence bias correction model tries to estimate the occurrence of a read with a surrounding sequence of each end from -8 to +12 nucleotides. A variable length Markov models were used to reduce the combinations of the 21 nucleotides, resulting in 774 parameters, as in \citep{Roberts2011}. In DICEseq, we estimate these parameters empirically from the genes with only one isoform. 

Empirically, we observed that correcting for biases did not significantly alter the results of our analyses, see Supplementary Table S2.

\subsection{Gene annotation, input datasets and processing}
Simulated reads in fastq format were generated from Spanki v0.5.0 \citep{Sturgill2013}. It is based on the human gene annotation and genome sequences which were downloaded from GENCODE with release 22. In addition to exclusively keeping protein coding genes, we further removed those genes that overlap with others or only have one isoform. Consequently, 90,759 isoforms from 11,426 genes were included for simulation. We randomly generated isoform ratios for each gene at 8 time points, with an assumption of either Gaussian process or first-order dynamics. Then, based on different overall coverages, the isoform ratios were multiplied to obtain the isoform specific reads-per-kilobase (RPK) value, which are used as the input values for Spanki simulator. 

4tU-seq data sets are available from the Gene Expression Omnibus (GEO; accession number GSE70378). The yeast gene annotation and genome sequences were downloaded from Ensembl with version R64-1-1, and all 309 intron-containing genes were included for analysis.

Circadian RNA-seq and microarray data sets on mouse liver were downloaded from GEO: GSE54652. The gene annotation and genome sequences were downloaded from GENCODE with release M6. Based on the annotation, we included 55,440 isoforms from 10,553 multiple-isoform, no-overlap, protein-coding genes. Processed microarray data \citep{Zhang2014}, which are based on Affymetrix MoGene 1.0 ST, were employed for validation of the isoform estimate from RNA-seq. The microarray probe ids were mapped to GENCODE ids by Ensembl BioMart, leaving 30534 isoforms from 9755 genes for study.

All above RNA-seq data sets were downloaded in fastq format, and first aligned to corresponding Genome sequences above via HISAT 0.1.5-beta \citep{Kim2015}, in paired-end mode with default setting.

\section{Results}

\subsection{Methods comparison using simulated reads}
In order to assess the performance of DICEseq, we compared it with three commonly used methods in their latest version: IsoEM v1.1.4, MISO v0.5.3, and Cufflinks v2.2.1. 
%IsoEM v1.1.4 \citep{Nicolae2011}, MISO v0.5.3 \citep{Katz2010}, and Cufflinks v2.2.1 \citep{Trapnell2010}. 
We also report results for a variant of DICEseq which ignores temporal correlations (DICE-sepa). Notice that DICE-sepa is essentially the same as MISO as a model, only differing in the estimation procedure and prior (collapsed M-H sampler and softmax of a Gaussian). Simulated reads for 11,462 human protein coding genes, accounting for a total of 90,759 distinct isoforms, were generated by Spanki v0.5.0 \citep{Sturgill2013}  with coverage from RPK of 50 to 1600 for 8 time points. We initially induced a temporal correlation between isoform proportions at different time points by enforcing the assumption of Gaussian process dynamics. All methods used paired-end reads, with the exception of MISO, which provided better performance in these experiments using single-end reads (see Supplementary Figure S2 and Table S3). We focus here on comparing the accuracy of the various methods; for a comparison of computational performance see Supplementary Figure S1.

We first studied the accuracy of each method at different coverage levels. We report average accuracy by computing the mean absolute error (MAE) between inferred isoform ratios and the truth from all the 90,759 isoforms of the 11,462 genes and 8 time points.  Figure 1A shows that all methods return accurate estimates, and that  the errors generally decrease with the increase of coverages. As expected, DICEseq is able to exploit effectively the temporal information, providing a significantly lower mean absolute error than the other methods, an advantage which is particularly marked at lower coverage. In a real RNA-seq time series experiment, many genes are likely to have relatively low coverage in at least one time point (see below our real data experiments), therefore the improved performance of DICEseq is likely to be important in quantifying isoforms for a substantial fraction of genes. A second, often very important, metric is the confidence intervals associated with the predictions. These can be useful when deciding e.g. which genes to include in downstream analyses as in \citep{Barrass2015}. We examined the average size of the confidence intervals for the three Bayesian methods DICEseq, MISO and Cufflinks as we vary the simulated coverage levels. As expected, confidence intervals shrink as we increase coverage for all three methods, however DICEseq clearly is able to provide more confident predictions at all coverage levels (Figure 1B). DICEseq is particularly strong at lower coverage; this is important, as often the confidence of an estimate is used to select genes which are further analysed \citep{Barrass2015}.

%\begin{figure}[!tpb]
\begin{figure*}[!tpb]
\centerline{\includegraphics[width=0.8\textwidth]{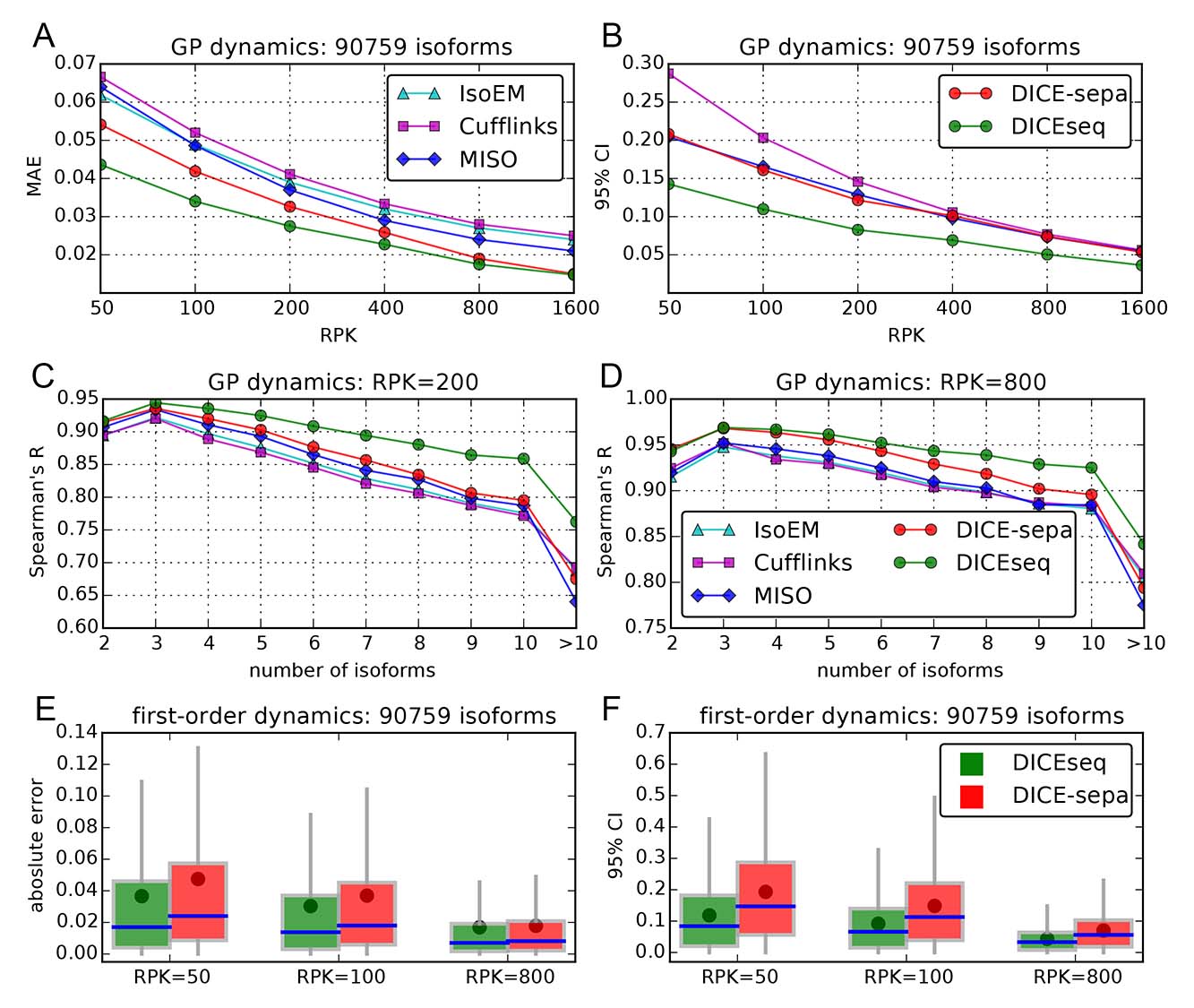}}
\caption{Comparison of accuracy between methods using simulated reads.
\textbf{(a)} Mean absolute error between estimated isoform proportion and the truth.  
\textbf{(b)} 95\% confidence interval of the estimates. 
\textbf{(c)} Influence of the number of isoforms on the estimates when RPK=200. 
\textbf{(d)} Influence of the number of isoforms on the estimates when RPK=800. The simulation is based on GP dynamics assumption for \textbf{(a-d)}.
\textbf{(e)} Boxplot of absolute error between estimated isoform proportion and the truth. 
\textbf{(f)} Boxplot of 95\% confidence interval of the estimates. The round dot is the mean. The simulation is based on first-order dynamics assumption for \textbf{(e-f)}.}
\end{figure*}

Thirdly, we investigate the influence of isoform number on the quality of the estimate at a specified coverage level. By selecting the genes with a specific number of isoform, Figure 1C (RPK=200) and 2D (RPK=800) both show that the rank correlation (Spearman's correlation) coefficient between the estimated isoform proportions and the truth generally decreases as the number of isoform increases. This is expected, because the presence of more isoforms reduces the number of uniquely assignable reads. Once again, we see that including temporal information can yield signficantly improved estimates, with DICEseq yielding an improvement in rank correlation of more than five percentage points for genes with many isoforms ($>$8).

Finally, we investigate the robustness of DICEseq to model mismatch. To do so, we generated time series data where the isoform proportions vary according to a first-order dynamical system (rather than a Gaussian process), a commonly used modelling hypothesis \citep{Eser2015}. Figure 1E-F clearly shows that incorporating temporal information yields a considerable improvement, even under model mismatch. This improvement is particularly marked at low coverages. Notice that the mean accuracy (represented by a dot in the box plots) is very similar to the one obtained under the GP assumption (Figure 1A).

In summary, the results of these simulation studies show that DICEseq can provide accurate reconstruction of isoform proportions, and can successfully leverage temporal information to provide more accurate and confident predictions at low coverage and for higher numbers of isoforms.

\subsection{Experiment design for time-series RNA-seq experiments}
Incorporating temporal information in the analysis of time series experiments is desirable in principle, because it provides experimentalists with a further direction for experimental design. Intuitively, resources can be invested in either improving the accuracy of each time point (by sequencing deeper), or by collecting more time points. This is an important trade-off, and it can only be achieved if the data is analysed jointly. To address these questions, we compared DICEseq versus DICE-sepa as we vary coverage levels and number of time points, by simulating reads as in the previous section (under GP assumption). In Figure 2A, we clearly see again that with the coverage increasing, all MAE decrease. In the joint model, the MAE largely decreases when more time points (i.e., 8) are used, especially for the case with low coverage. 
\begin{figure*}[!pb]
\centerline{\includegraphics[width=0.8\textwidth]{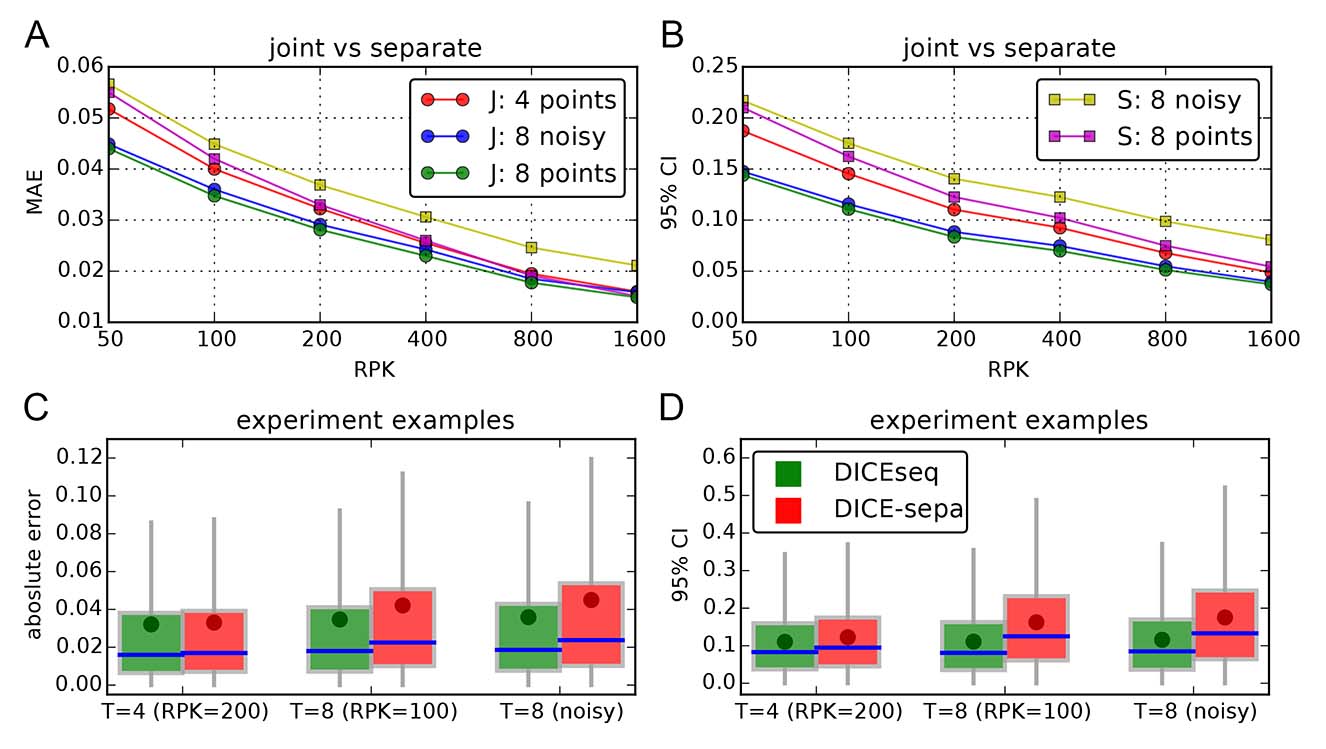}}
\caption{Comparison between experiment design on time points and coverages.
\textbf{(a)} Mean absolute error between estimated isoform proportion and the truth for different experiments. "S" and "J" means DICEseq separate and joint mode, respectively. The "noisy" means the RPK=25 at the 5th point. All simulation here is based on GP dynamics assumption. 
\textbf{(b)} 95\% confidence interval of the estimates. 
\textbf{(c)} Boxplot of absolute error between estimated isoform proportion and the truth for three example experiments. The round dot is the mean. "T=4" and "T=8" means the 4 and 8 number of time points. The "noisy" example was also conducted at RPK=100.
\textbf{(d)} Boxplot of 95\% confidence interval of the estimates.}
\end{figure*}

These results highlight the importance of the analysis method for experimental design: while with the non-temporal model DICE-sepa increasing coverage is the only way to improve accuracy, methods that incorporate temporal information can benefit both from an increase in coverage and an increase in sampling frequency. Broadly speaking, we see that a doubling of the sampling frequency is roughly equivalent to a doubling of the sequencing depth, with the obvious advantage that a finer temporal information is provided. Figure 2C and 3D show an example of this trade-off:  4 time points and higher coverage of $RPK=200$ give indistinguishable results for the joint model to 8 time points and lower coverage of $RPK=100$ (first two pairs in Figure 2C/D). 

Another potential advantage of incorporating temporal information is to improve robustness of the estimation against noise/ low coverage at some time points. This aspect is particularly important as of course coverage level for a particular gene is largely determined by the gene's expression level, therefore genes with a large dynamic range of expressions during the time series will necessarily have some time points with low coverage. To simulate this situation, we generated time series with very low coverage ($RPK=25$) in the 5th time point. From  the {"noisy" case in} Figure 2, we could see that the joint model dramatically reduces the variation compared to the separated model. Thus, incorporating time information in the joint model leads to a more robust estimation, facilitating isoform estimation for genes with dynamic expression levels and providing a possibility to combine low coverage with high coverage time points for time series libraries.

\subsection{RNA splicing dynamics with 4tU-seq data}\label{4tUanalysis}
Recently, biotin labelling combined with RNA-seq has become an important tool to study the kinetics of RNA transcription and splicing with high temporal resolution \citep{Windhager2012,Veloso2014,Fuchs2014}. These experiments naturally produce RNA-seq data sets with high temporal resolution; furthermore, at very early time points, labelled RNA may be of low abundance, resulting in high uncertainty estimates. Here we use a recent data set with high temporal resolution to probe the suitability of DICEseq as an analysis tool for biotin labelled RNA-seq; the data was produced by our collaborators in the Beggs and Granneman labs at the Wellcome Trust Centre for Cell Biology in Edinburgh \citep{Barrass2015}. The data set consists of approximately 50M mapped reads; roughly 50\% of genes have a coverage of RPK$<$120 in at least one time point. 
\begin{figure*}[!pb]
\centerline{\includegraphics[width=0.8\textwidth]{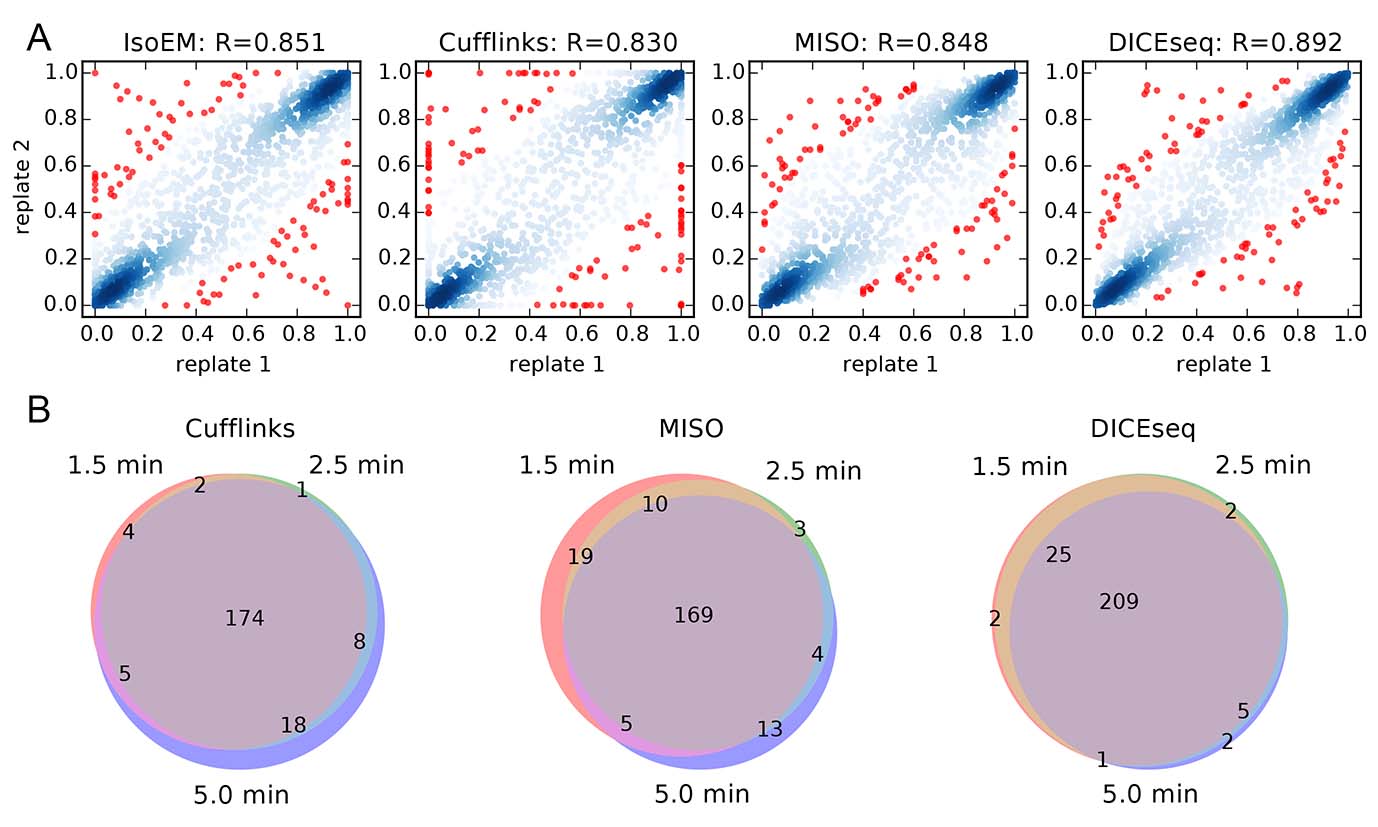}}
\caption{Analysis of time series 4tU-seq data.
\textbf{(a)} The Pearson's correlation between two replicates.
\textbf{(b)} The number of genes whose 95\% confidence interval $<$ 0.3.}
\end{figure*}

To assess accuracy of our method, we compare the correlation between two replicates for 309 intron-containing genes at 1.5, 2.5 and 5.0 minute. Figure 3A shows that IsoEM, Cufflinks and MISO all result in a good correlation between replicates, with Pearson's correlation coefficient varying between 0.83 and 0.85; DICEseq further improves with a Pearson's correlation coefficient of 0.892, outperforming by between 4 and 6 percentage points existing methods. The improvement is particularly marked if we consider the lowest expressed genes: on the lower third of the expression range, DICEseq still obtains a Pearson correlation of 0.834, while the other methods achieve much lower correlations, ranging from 0.657 (Cufflinks) to 0.738 (IsoEM). This is remarkable since, as there are only three time points, the improvement obtained by taking temporal information into account could be expected to be limited. Notice in particular that, while IsoEM and particularly Cufflinks sometimes give deterministic estimates in one replicate but not on the other (red points on the boundaries of the square in Figure 3A), this problem does not occur with DICEseq, presumably due to the stronger regularisation enforced by the temporal correlations.

To further explore the usefulness of DICEseq, we consider the confidence intervals reported by the various methods. Isoform quantification methods are often used as an initial step in kinetic analyses of individual transcripts; in order to reduce false positives, genes with unreliable isoform estimates (as determined by thresholding on the confidence intervals) are discarded. When quantifying isoforms in isolation, some genes are then discarded just because one of the time points have lower expression level. Therefore, we computed the number of transcripts that pass a frequently used threshold (95\%CI$<0.3$) for further analysis \citep{Barrass2015}. Figure 3B illustrates the results, showing that at all time points between  10\% and 20\% more genes are retained using a joint analysis, compared to methods that analyse data points in isolation.
%\footnote{Note that Cufflinks returns confidence interval as RPK score, rather than isoform proportion, so the interpretation could be slightly different.}.

To summarise, our results on a real yeast kinetic data set confirm that DICEseq yields significantly more reproducible and confident results than existing state-of-the-art methods, highlighting the value of incorporating temporal information in the analysis of time series real data.

\subsection{Circadian dynamics of alternative splicing}
As a second real-data example, we turned to a recent data set investigating circadian control of gene expression in mouse. Due to the day-night oscillations, many biological processes, including gene expression, show circadian rhythms. %\citep{Neill2011}. 
Recently, Zhang \emph{et al} \citep{Zhang2014} systematically studied circadian gene expression on 12 mouse tissues using high-temporal resolution microarrays and RNA-seq, and found that 43\% protein coding genes oscillate in at least on one of the 12 tissues; here we focus on data from liver. The RNA-seq here has a comparably low time resolution, as eight time points were collected over a period of 48 hours; we expect therefore that the advantages of incorporating time information may be less pronounced in this scenario. In total, there are between 67M and 105M uniquely mapped reads in each experiment; on average of 8 time points, 50\% of genes have all isoforms with RPK$<$70; 75\% of genes have all isoform with RPK$<$400.
\begin{figure*}[!tpb]
\centerline{\includegraphics[width=0.8\textwidth]{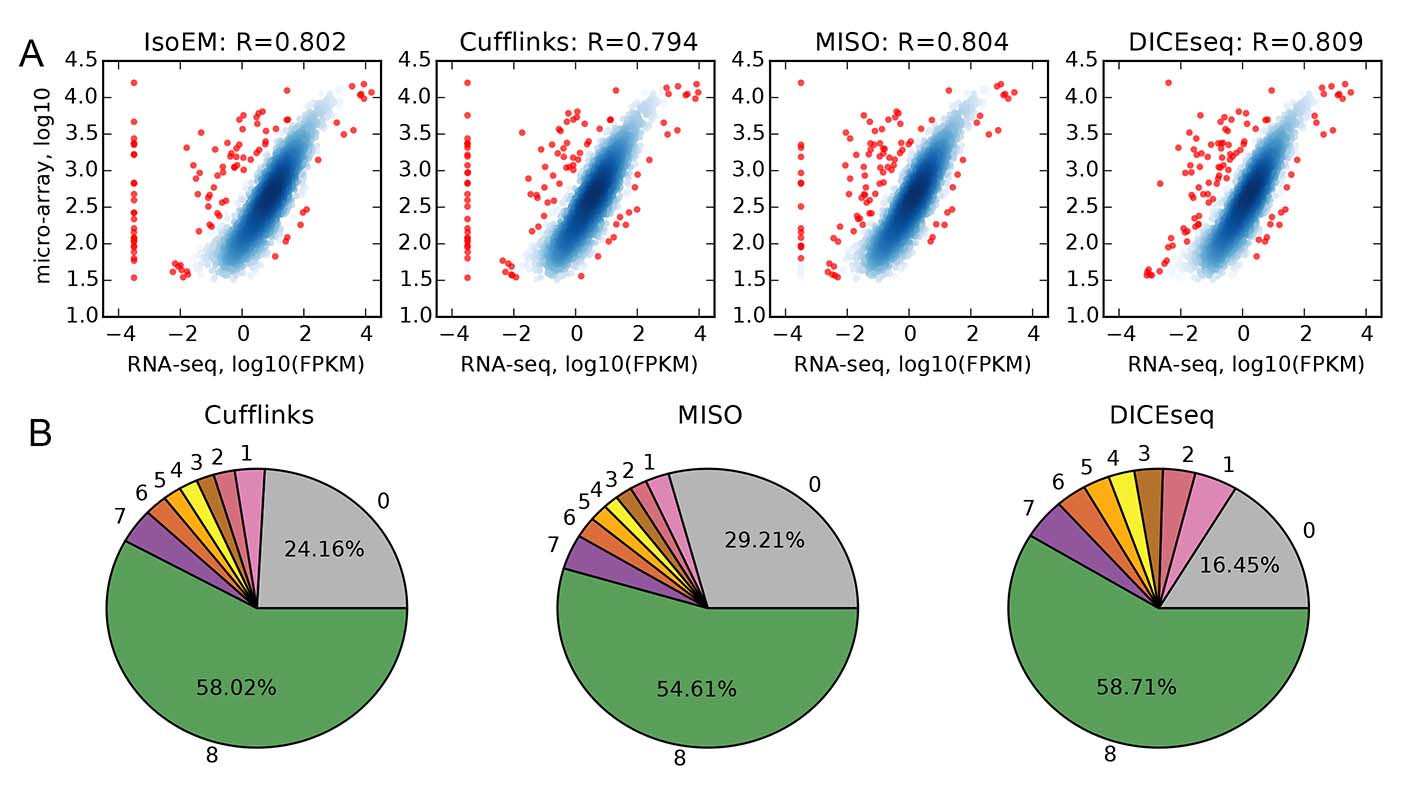}}
\caption{Analysis of circadian time series data.
\textbf{(a)} The Spearman's correlation between the measurement of RNA-seq and microarray.
\textbf{(b)} The proportion of genes whose 95\% confidence interval $<$ 0.3 in a certain number of time points (index on the external side of the circle).}
\end{figure*}

To assess the performance of the various methods, we used the microarray data set to validate the isoform estimates from RNA-seq. Unfortunately, only about one hundred microarray probes map to a unique annotated isoform (out of 30,534 annotated isoforms which map to at least on microarray probe); in other words, most microarray probes map to multiple isoform within a gene. Thus, we used the estimated isoform proportions, together with the total numbers of reads mapped to each gene, to predict the expression level (as FPKM) of each isoform covered by at least one microarray probe, and then compared the resulting estimate from RNA-seq with the microarray measurement. As the two methods have different response ranges/ dynamics, we used Spearman's rank correlation as a metric.
In Figure 4A, we see that the estimates obtained from RNA-seq using all methods have a very high correlation with the direct measurements from the microarrays. Still, DICEseq shows a slightly improved correlation; in particular, very low expressed isoforms (outliers in the left end of the plot) show a much better quantification with DICEseq than with the other methods, probably due to the sharing of the temporal information. 

We further measured 95\% confidence intervals (CI) of all the 55,440 isoforms at the 8 time points, and quantified the fraction of isoform quantifications that pass the threshold 95\%CI$<$0.3. In Figure 4B we see that all Bayesian methods (Cufflinks, MISO and DICEseq) give confident estimates for between 50 and 60 \% of isoforms at all time points. Once again, DICEseq estimates are more confident, thanks to the value of temporal information sharing at low coverages, even though the advantage is more modest in this data set.

To summarise, our results in this low-frequency RNA-seq time series data set show that even in this case DICEseq produces quantitatively better estimates of isoform ratios, even though the value of sharing temporal information is more limited here due to the weaker correlations between time points. 
\begin{table}[h!]
\caption{Robust performance of DICEseq in lower or medium coverage. "All" means all annotated genes; "1/3 low" and "1/3 mid" respectively mean lowest and medium 1/3 genes in coverage.
In the 4tU-seq experiment, the scores are the Pearson's correlation coefficient between two replicates. In the circadian experiment, the scores are the Spearman's correlation coefficient between the measurement of RNA-seq and microarray.}
\centerline{
      \begin{tabular}{lcccc}
        \hline
                    & IsoEM  &Cufflinks &MISO  &DICEseq \\ \hline
 4tU-seq, all       & 0.851  &0.830     &0.848 &0.892   \\
 4tU-seq, 1/3 low   & 0.738  &0.657     &0.701 &0.834   \\ \hline
 circadian, all     & 0.802  &0.794     &0.804 &0.809   \\
 circadian, 1/3 mid & 0.568  &0.555     &0.577 &0.589   \\ \hline
      \end{tabular}
}
\end{table}

\section{Discussion}
The advent of RNA-seq technologies has revolutionised the study of mRNA splicing, and provided a powerful stimulus for the development of computational biology methods \citep{Katz2010,Trapnell2010,Nicolae2011,Glaus2012}. Recent years have seen a more wide-spread use of RNA-seq technology for the analysis of dynamical biological processes, resulting in a marked increase of biological studies adopting RNA-seq within a time series experimental design. In this article, we presented DICEseq, the first method to jointly estimate the dynamics of the splicing isoform proportions from time series RNA-seq data. A comparison of DICEseq to a selection of popular state-of-the-art methods shows that DICEseq has excellent accuracy and good computational performance; in particular, DICEseq can effectively pool information across time points to improve isoform quantification at low coverages, giving more accurate and confident predictions. Our analysis also points to the importance of coverage versus temporal sampling trade-offs in designing dynamic RNA-seq experiments; while our analysis focussed on time series experiments, we expect similar considerations to hold for other structured designs, such as dose response experiments. In this light, the use of methods which can capture structural information, such as DICEseq, may lead to a rethink of biological experimental designs for a broad class of experiments. Our application to two diverse biological data sets shows that DICEseq can be an effective tool on real biological investigations, leading to improved performance and more reproducible results.

Methodologically, DICEseq builds on a fertile line of research using GPs to model transcriptional dynamics. GPs have been used to study the dynamical behaviour of gene expression in various contexts, from transcriptional regulation \citep{Lawrence:modelling06} to identifying the time intervals of differential expression with time series microarray data \citep{Stegle2010}. More recently, \"{A}ij\"{o} \emph{et al} used a latent GP with negative binomial observation noise to study the profiles of gene expression during Th17 cell differentiation with time course RNA-seq \citep{Aijo2014}. To our knowledge, this is the first time GPs have been proposed within the context of isoform estimation.

While we believe DICEseq offers a valuable new tool for the analysis of dynamic RNA-seq data, it also opens several novel lines of investigation. Firstly, the Gaussian process prior, which is based on a general regression, could be extended to more general dynamic splicing modeling, e.g., a first-order linear dynamic system for RNA splicing kinetics, and an oscillatory system for circadian or cell-cycle studies. All of these could be incorporated in a straightforward way as parametric mean functions in a GP framework, however it would also be of interest to explicitly model the noise correlations they induce. DICEseq could be useful in elucidating RNA processing from biotin labelled RNA-seq, as attempted e.g. in de Pretis \emph{et al} \citep{Pretis07052015}. More generally, DICEseq could provide a flexible Bayesian framework for explaining RNA-seq data from other observations, and aid studies attempting to link splicing with other genetic and epigenetic factors. %\citep{Ye2013,Gelfman2013,Curado2015}.

%%%%%%%%%%%%%%%%%%%%%%%%%%%%%%%%%%%%%%%%%%%%%%%%%%%%%%%%%%%%%%%%%%%%%%%%%%%%%%%%%%%%%
%
%     please remove the " % " symbol from \centerline{\includegraphics{fig01.jpg}}
%     as it may ignore the figures.
%
%%%%%%%%%%%%%%%%%%%%%%%%%%%%%%%%%%%%%%%%%%%%%%%%%%%%%%%%%%%%%%%%%%%%%%%%%%%%%%%%%%%%%%

\section*{Acknowledgements}
We would like to thank Prof. Jean D Beggs, Dr. Sander Granneman, Dr. Jane E A Reid, Dr. J David Barrass, Dr Edward Wallace and Dr. Yichuan Zhang for fruitful discussions. G.S. acknowledges support from the European Research Council under grant MLCS306999. Y.H. is supported by the University of Edinburgh through a Principal Career Development scholarship.\vspace*{-12pt}

\bibliographystyle{genres}
\bibliography{diceseq.bib}       % Bibliography file (usually '*.bib' )

\begin{thebibliography}{}

\bibitem[{\"A}ij{\"o} et~al., 2014]{Aijo2014}
{\"A}ij{\"o}, T., Butty, V., Chen, Z., Salo, V., Tripathi, S., Burge, C.~B.,
  Lahesmaa, R., and L{\"a}hdesm{\"a}ki, H., 2014.
\newblock {Methods for time series analysis of RNA-seq data with application to
  human Th17 cell differentiation}.
\newblock {\em Bioinformatics}, \textbf{30}(12):i113--i120.

\bibitem[Bar-Joseph et~al., 2012]{Bar2012}
Bar-Joseph, Z., Gitter, A., and Simon, I., 2012.
\newblock {Studying and modelling dynamic biological processes using
  time-series gene expression data}.
\newblock {\em Nature Reviews Genetics}, \textbf{13}(8):552--564.

\bibitem[Barrass et~al., 2015]{Barrass2015}
Barrass, J.~D., Reid, J.~E., Huang, Y., Hector, R.~D., Sanguinetti, G., Beggs,
  J.~D., and Granneman, S., 2015.
\newblock {Transcriptome-wide RNA processing kinetics revealed using extremely
  short 4tU labeling}.
\newblock {\em Genome Biology}, \textbf{16}(1):1--17.

\bibitem[Blencowe, 2006]{Blencowe2006}
Blencowe, B.~J., 2006.
\newblock {Alternative splicing: new insights from global analyses}.
\newblock {\em Cell}, \textbf{126}(1):37--47.

\bibitem[de~Pretis et~al., 2015]{Pretis07052015}
de~Pretis, S., Kress, T., Morelli, M.~J., Melloni, G.~E., Riva, L., Amati, B.,
  and Pelizzola, M., 2015.
\newblock {INSPEcT: a Computational Tool to Infer mRNA Synthesis, Processing
  and Degradation Dynamics from RNA-and 4sU-seq Time Course Experiments}.
\newblock {\em Bioinformatics}, \textbf{}:btv288.

\bibitem[Eser et~al., 2015]{Eser2015}
Eser, P., Wachutka, L., Maier, K.~C., Demel, C., Boroni, M., Iyer, S., Cramer,
  P., and Gagneur, J., 2015.
\newblock {Determinants of RNA metabolism in the Schizosaccharomyces pombe
  genome}.
\newblock {\em bioRxiv}, \textbf{}:025585.

\bibitem[Fuchs et~al., 2014]{Fuchs2014}
Fuchs, G., Voichek, Y., Benjamin, S., Gilad, S., Amit, I., and Oren, M., 2014.
\newblock {4sUDRB-seq: measuring genomewide transcriptional elongation rates
  and initiation frequencies within cells}.
\newblock {\em Genome Biology}, \textbf{15}(5):R69.

\bibitem[Glaus et~al., 2012]{Glaus2012}
Glaus, P., Honkela, A., and Rattray, M., 2012.
\newblock {Identifying differentially expressed transcripts from RNA-seq data
  with biological variation}.
\newblock {\em Bioinformatics}, \textbf{28}(13):1721--1728.

\bibitem[Graveley, 2001]{Graveley2001}
Graveley, B.~R., 2001.
\newblock {Alternative splicing: increasing diversity in the proteomic world}.
\newblock {\em TRENDS in Genetics}, \textbf{17}(2):100--107.

\bibitem[Honkela et~al., 2015]{Honkela2015}
Honkela, A., Peltonen, J., Topa, H., Charapitsa, I., Matarese, F., Grote, K.,
  Stunnenberg, H.~G., Reid, G., Lawrence, N.~D., and Rattray, M.,
  \emph{et~al.}, 2015.
\newblock {Genome-wide modeling of transcription kinetics reveals patterns of
  RNA production delays}.
\newblock {\em Proceedings of the National Academy of Sciences},
  \textbf{112}(42):13115--13120.

\bibitem[Kanitz et~al., 2015]{Kanitz2015}
Kanitz, A., Gypas, F., Gruber, A.~J., Gruber, A.~R., Martin, G., and Zavolan,
  M., 2015.
\newblock {Comparative assessment of methods for the computational inference of
  transcript isoform abundance from RNA-seq data}.
\newblock {\em Genome Biology}, \textbf{16}(1):1--26.

\bibitem[Katz et~al., 2010]{Katz2010}
Katz, Y., Wang, E.~T., Airoldi, E.~M., and Burge, C.~B., 2010.
\newblock {Analysis and design of RNA sequencing experiments for identifying
  isoform regulation}.
\newblock {\em Nature Methods}, \textbf{7}(12):1009--1015.

\bibitem[Kim et~al., 2015]{Kim2015}
Kim, D., Langmead, B., and Salzberg, S.~L., 2015.
\newblock {HISAT: a fast spliced aligner with low memory requirements}.
\newblock {\em Nature Methods}, \textbf{12}(4):357--360.

\bibitem[Lawrence et~al., 2006]{Lawrence:modelling06}
Lawrence, N.~D., Sanguinetti, G., and Rattray, M., 2006.
\newblock {Modelling transcriptional regulation using {G}aussian processes}.
\newblock In {\em Advances in Neural Information Processing Systems}, pages
  785--792.

\bibitem[Nicolae et~al., 2011]{Nicolae2011}
Nicolae, M., Mangul, S., Mandoiu, I.~I., and Zelikovsky, A., 2011.
\newblock {Estimation of alternative splicing isoform frequencies from RNA-Seq
  data}.
\newblock {\em Algorithms for Molecular Biology}, \textbf{6}(1):9.

\bibitem[Rasmussen and Williams, 2006]{Rasmussen2006}
Rasmussen, C.~E. and Williams, C. K.~I., 2006.
\newblock {\em {Gaussian processes for machine learning}}.
\newblock MIT Press, Cambridge, MA, USA.

\bibitem[Roberts et~al., 2011]{Roberts2011}
Roberts, A., Trapnell, C., Donaghey, J., Rinn, J.~L., Pachter, L., et~al.,
  2011.
\newblock {Improving RNA-Seq expression estimates by correcting for fragment
  bias}.
\newblock {\em Genome Biology}, \textbf{12}(3):R22.

\bibitem[Scotti and Swanson, 2016]{Scotti2016}
Scotti, M.~M. and Swanson, M.~S., 2016.
\newblock {RNA mis-splicing in disease}.
\newblock {\em Nature Reviews Genetics}, \textbf{17}(1):19--32.

\bibitem[Stegle et~al., 2010]{Stegle2010}
Stegle, O., Denby, K.~J., Cooke, E.~J., Wild, D.~L., Ghahramani, Z., and
  Borgwardt, K.~M., 2010.
\newblock {A robust Bayesian two-sample test for detecting intervals of
  differential gene expression in microarray time series}.
\newblock {\em Journal of Computational Biology}, \textbf{17}(3):355--367.

\bibitem[Sturgill et~al., 2013]{Sturgill2013}
Sturgill, D., Malone, J.~H., Sun, X., Smith, H.~E., Rabinow, L., Samson, M.-L.,
  and Oliver, B., 2013.
\newblock {Design of RNA splicing analysis null models for post hoc filtering
  of Drosophila head RNA-Seq data with the splicing analysis kit (Spanki)}.
\newblock {\em BMC bioinformatics}, \textbf{14}(1):320.

\bibitem[Trapnell et~al., 2010]{Trapnell2010}
Trapnell, C., Williams, B.~A., Pertea, G., Mortazavi, A., Kwan, G., van Baren,
  M.~J., Salzberg, S.~L., Wold, B.~J., and Pachter, L., 2010.
\newblock {Transcript assembly and quantification by RNA-Seq reveals
  unannotated transcripts and isoform switching during cell differentiation}.
\newblock {\em Nature Biotechnology}, \textbf{28}(5):511--515.

\bibitem[Tuomela et~al., 2012]{Tuomela2012}
Tuomela, S., Salo, V., Tripathi, S.~K., Chen, Z., Laurila, K., Gupta, B.,
  {\"A}ij{\"o}, T., Oikari, L., Stockinger, B., L{\"a}hdesm{\"a}ki, H.,
  \emph{et~al.}, 2012.
\newblock {Identification of early gene expression changes during human Th17
  cell differentiation}.
\newblock {\em Blood}, \textbf{119}(23):e151--e160.

\bibitem[Veloso et~al., 2014]{Veloso2014}
Veloso, A., Kirkconnell, K.~S., Magnuson, B., Biewen, B., Paulsen, M.~T.,
  Wilson, T.~E., and Ljungman, M., 2014.
\newblock {Rate of elongation by RNA polymerase II is associated with specific
  gene features and epigenetic modifications}.
\newblock {\em Genome Research}, \textbf{24}(6):896--905.

\bibitem[Wang et~al., 2008]{Wang2008}
Wang, E.~T., Sandberg, R., Luo, S., Khrebtukova, I., Zhang, L., Mayr, C.,
  Kingsmore, S.~F., Schroth, G.~P., and Burge, C.~B., 2008.
\newblock {Alternative isoform regulation in human tissue transcriptomes}.
\newblock {\em Nature}, \textbf{456}(7221):470--476.

\bibitem[Wang et~al., 2009]{Wang2009}
Wang, Z., Gerstein, M., and Snyder, M., 2009.
\newblock {RNA-Seq: a revolutionary tool for transcriptomics}.
\newblock {\em Nature Reviews Genetics}, \textbf{10}(1):57--63.

\bibitem[Windhager et~al., 2012]{Windhager2012}
Windhager, L., Bonfert, T., Burger, K., Ruzsics, Z., Krebs, S., Kaufmann, S.,
  Malterer, G., L'Hernault, A., Schilhabel, M., Schreiber, S., \emph{et~al.},
  2012.
\newblock {Ultrashort and progressive 4sU-tagging reveals key characteristics
  of RNA processing at nucleotide resolution}.
\newblock {\em Genome Research}, \textbf{22}(10):2031--2042.

\bibitem[Zhang et~al., 2014]{Zhang2014}
Zhang, R., Lahens, N.~F., Ballance, H.~I., Hughes, M.~E., and Hogenesch, J.~B.,
  2014.
\newblock {A circadian gene expression atlas in mammals: Implications for
  biology and medicine}.
\newblock {\em Proceedings of the National Academy of Sciences},
  \textbf{111}(45):16219--16224.

\end{thebibliography}

\newpage
\begin{appendices}
\renewcommand{\thefigure}{S\arabic{figure}}
\renewcommand{\thetable}{S\arabic{table}} 

\section{Computation performance}
We first investigate the speed of these methods (IsoEM, Cufflinks, MISO, DICEseq, and its separate mode DICE-sepa) at different coverages. Here, 4 out 64 CPU cores were parallel used for Cufflinks, MISO, DICEseq, and DICE-sepa. IsoEM was used as default in multiple cores setting. From supplementary Figure S1, we see that IsoEM is the fastest method, followed by Cufflinks. Though MISO, DICEseq and DICE-sepa are much slower than IsoEM and Cufflinks, they still finish the 8 time points within a reasonable period. Also, we noticed that the running time for both DICEseq and DICE-sepa increases much slower along the coverage than other methods. Therefore, we further tested the running time for different number of genes in the annotation file on a single ENCODE library. And we found that the running time for both MISO and DICEseq is almost linearly correlated with the number of genes for estimate. Notably, if there are only a fewer hundreds genes for study (e.g., total $\sim$300 intron-containing genes in yeast), DICEseq could finish the job within 10 minutes for a single time point.

\section{Supplementary Figures}   
\setcounter{figure}{0}  
\begin{figure*}[h!]
\begin{center}
\includegraphics[width=0.9\textwidth]{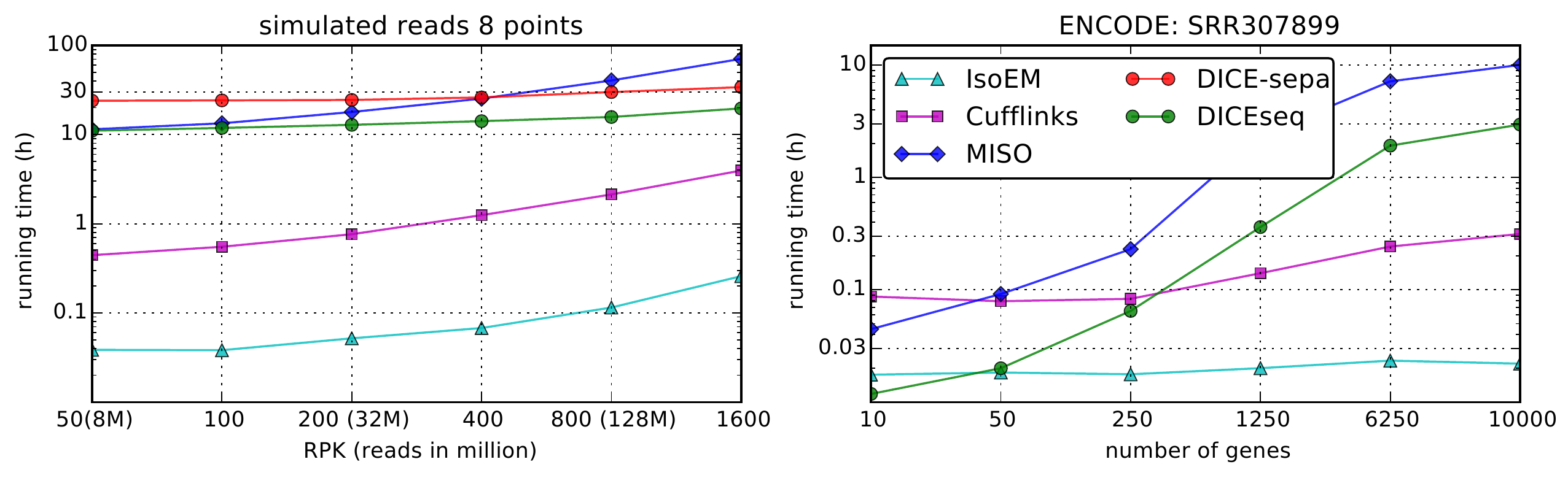}
\end{center}
\caption{Comparison of the running time between methods.
Running time on simulated libraries with different coverages (left panel). 
Running time on an ENCODE library for different number of genes (right panel).}
\end{figure*}

\begin{figure*}[h!]
\begin{center}
\includegraphics[width=0.9\textwidth]{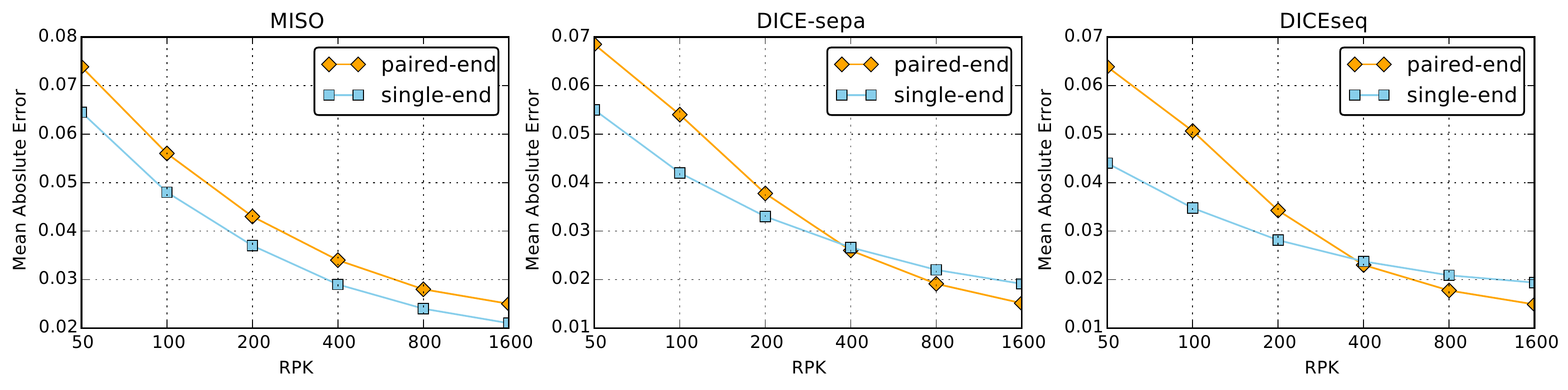}
\end{center}
\caption{Comparison between using single-end and paired-end reads for MISO (left panel), DICE-sepa (middle panel) and DICEseq (right panel) on simulated reads, by measuring the mean absolute error between the estimates and the truth.}
\end{figure*}

\section{Supplementary Tables}
\setcounter{table}{0}
\begin{table}[h!]
\caption{Comparison of $\theta_2$ by measuring the mean absolute errors between inference and truth. A total of 1,547 isoforms from 198 random human genes were used in the simulation. In the first row, we show the percentage of the length that $\theta_2$ covers. GP means the assumption of Gaussian process dynamics, and FO means the first-order dynamics. RPK100 and RPK800 mean different sequencing coverages.}
\centering
      \begin{tabular}{cccccccc}
        \hline
covers of $\theta_2$    &0\%    &4.5\%  &15\%   &33\%   &65\%   &100\%  &200\%   \\ \hline
   RPK100, GP           &0.118  &0.059  &0.055  &0.054  &0.060  &0.066  &0.124   \\
   RPK800, GP           &0.101  &0.020  &0.019  &0.017  &0.024  &0.032  &0.113   \\ \hline
   RPK100, FO           &0.076  &0.058  &0.055  &0.052  &0.050  &0.049  &0.123   \\
   RPK800, FO           &0.042  &0.021  &0.019  &0.017  &0.015  &0.015  &0.116   \\ \hline
      \end{tabular}
\end{table}
\vspace{-1em}

\begin{table}[h!]
\caption{Comparison between uniform and biased distribution of reads.
In the 4tU-seq experiment, the scores are the Pearson's correlation coefficient between two replicates. In the circadian experiment, the scores are the Spearman's correlation coefficient between the measurements of RNA-seq and microarray. MISO does not support bias correction, and IsoEM failed to return results when correcting bias for replicate 2 in 4tU-seq experiment.}
\centering
      \begin{tabular}{ccccc}
        \hline
                  & IsoEM  &Cufflinks &MISO   &DICEseq \\ \hline
 4tU-seq, unif    & 0.851  &0.830     &0.843  &0.892   \\
 4tU-seq, bias    & X      &0.839     &X      &0.897   \\ \hline
 circadian, unif  & 0.802  &0.794     &0.804  &0.809   \\
 circadian, bias  & 0.803  &0.796     &X      &0.807   \\ \hline
      \end{tabular}
\end{table}
\vspace{-1em}

\begin{table}[h!]
\caption{Comparison between using single-end (SE) and paired-end (PE) reads in circadian experiment. The second and third rows are the the Spearman's correlation coefficient between the measurement of RNA-seq and microarray. The fourth and fifth rows are the number of genes which passed the threshold of 95\%$<$0.3 at all 8 time points.}
\centering
      \begin{tabular}{cccc}
        \hline
                     &MISO  &DICEseq \\ \hline
   Spearmans' R, SE  &0.805 &0.807   \\
   Spearmans' R, PE  &0.804 &0.809   \\ \hline
   N(95\%$<$0.3), SE &34213 &35424   \\
   N(95\%$<$0.3), PE &30275 &32551   \\ \hline
      \end{tabular}
\end{table}

\end{appendices}

\end{document}